\documentclass[11pt]{article}
\usepackage{amsmath}
\usepackage{amsfonts}
\usepackage{amssymb}
\usepackage{graphicx}

\def\bea{\begin{eqnarray}}
\def\eea{\end{eqnarray}}

\begin{document}
\begin{center}
\LARGE { \bf Hidden Conformal Symmetry of Extremal Kerr-Bolt Spacetimes
  }
\end{center}
\begin{center}
{\bf M. R. Setare\footnote{rezakord@ipm.ir} \\  V. Kamali\footnote{vkamali1362@gmail.com}}\\
 {\ Department of Science of Bijar, University of  Kurdistan  \\
Bijar, IRAN.}
 \\
 \end{center}
\vskip 3cm

\begin{abstract}
We show that  extremal Kerr-Bolt spacetimes have a hidden conformal symmetry. In this regard, we consider the wave equation of a massless scalar field propagating in extremal Kerr-Bolt spacetimes and find in the "near region", the wave equation in extremal limit can be written in terms of the $SL(2,R)$ quadratic Casimir.

Moreover, we obtain the microscopic entropy of the extremal Kerr-Bolt spacetimes also we calculate the correlation function of a near-region scalar field  and find perfect agreement with the dual 2D CFT.
\\

\end{abstract}

\newpage

\section{Introduction}
In the context of proposed Kerr/CFT correspondence \cite{stro}, the microscopic entropy of four-dimensional extremal Kerr black hole is calculated by studying the dual chiral conformal field theory associated with the diffeomorphisms of near horizon geometry of the Kerr black hole. The isometry of the near horizon geometry of extremal Kerr is given by $SL(2,R) \times U(1)$. The authors of \cite{stro} considered the enhancement of the $U(1)$
isometry to the Virasoro algebra by using the asymptotic symmetry. After that an another Virasoro algebra was
obtained by extending the $SL(2,R)$ part of the isometry \cite{2}. The former Virasoro
algebra is referred to as that of the left movers while the latter as that of the right
movers. For the other works done on Kerr
/CFT correspondence see \cite{KCFT}. Subsequently, the Kerr/CFT correspondence was extended to the case of near-extremal black holes \cite{21} (see also \cite{3}). The main progresses are made essentially on the extremal and near extremal limits in which the black hole near horizon geometries consist a certain AdS structure and the central charges of dual CFT can be obtained by analyzing
the asymptotic symmetry following the method in \cite{bh} or by calculating the boundary stress tensor of the $2D$ effective action \cite{4}.
 Recently, Castro, Maloney and Strominger \cite{cms} have given evidence that the physics of non-extremal Kerr black holes might be captured by a conformal field theory. The authors have discussed that the existence of conformal invariance in a near horizon geometry is not a necessary condition, instead the existence of a local conformal invariance in the solution space of the wave equation for the propagating field is sufficient to ensure a dual CFT description. The scalar Laplacian
in the low frequency limit could be written as the $SL(2,R)$ quadratic Casimir, showing a
hidden $SL(2,R) × SL(2,R)$ symmetries. Even though these $SL(2,R)$ symmetries are not
global defined and are broken by the angular identification $\varphi\sim\varphi+2\pi$, they act on the
solution space, determine the form of the scattering amplitudes. The existence of this hidden conformal symmetry is essential
to set up a CFT dual to non-extremal Kerr black hole (see \cite{recent} for recent works in this topic). More  recently the authors of \cite{6} have found that the Laplacian of the scalar field in many extremal black holes also could be written in terms
of the $SL(2,R)$ quadratic Casimir. This suggests that there exist dual CFT descriptions
of these black holes. They have introduced a new set of conformal coordinates and have
computed the corresponding $SL(2,R)$ quadratic Casimir.\\
The spacetimes with NUT twist have been studied extensively in regard to their conserved charges, maximal mass conjecture and D-bound in \cite{TN1}.
These spacetimes have conical and Dirac-Misner singularities that should be removed by identifications of coordinates in the metric. Following \cite{6} we
consider the four-dimensional extremal Kerr-Bolt spacetimes and show they have a hidden conformal
symmetry. We consider the wave equation of a massless scalar field in the
background of extremal Kerr-Bolt spacetimes and  in the "near region", we obtain the radial part of wave equation. Then in a new conformal coordinates we obtain the quadratic Casimir of $ SL(2;R)$. The crucial observation is that these Casimir, when written in term of coordinates
$\varphi$, $t$ and $r$ reduces to the radial equation of massless scalar field. After that we obtain the
macroscopic entropy of the  extremal Kerr-Bolt spacetimes. Moreover we find the absorption cross
section of a near-region scalar field matches to the microscopic cross section in dual CFT.
\section{Extremal Kerr-Bolt black hole }
In this section, we give a brief review of the Kerr-Bolt black hole  \cite{8}. Its metric takes the following form in Boyer-Lindquist type coordinates

\begin{eqnarray}\label{1}
ds^2=-\frac{\Delta_r}{\Xi^2\rho^2}[dt+(2n\cos\theta -a\sin^2\theta)d\varphi]^2~~~~~~~~~~~~~~~~~~~~~~~~~~~~~~~~~~~~~\\
\nonumber+\frac{\Delta_{\theta}\sin^2\theta}{\Xi^2\rho^2}[adt-(r^2+n^2+a^2)d\varphi]^2~~~~~~~~~~~~~~~~~~~~~~~~~~~~~~~~~~~~~~~~~~\\
\nonumber
+\frac{\rho^2dr^2}{\Delta_r}+\frac{\rho^2d\theta^2}{\Delta_\theta}~~~~~~~~~~~~~~~~~~~~~~~~~~~~~~~~~~~~~~~~~~~~~~~~~~~~~~~~~~~~~~~~~~~~
  \end{eqnarray}
  where

\begin{eqnarray}\label{2}
\rho^2=r^2+(n+a\cos\theta)^2~~~~~~~~~~~~~~~~~~~~~~~~~~~~~~~~~~~~~~~~~~~~~~~~\\
\nonumber
\Delta_r=-\frac{r^2(r^2+6n^2+a^2)}{l^2}+r^2-2mr-\frac{(3n^2-l^2)(a^2-n^2)}{l^2}\\
\nonumber
\Delta_{\theta}=1+\frac{a\cos\theta(4n+a\cos\theta)}{l^2}~~~~~~~~~~~~~~~~~~~~~~~~~~~~~~~~~~~~~~~\\
\nonumber
\Xi=1+\frac{a^2}{l^2}~~~~~~~~~~~~~~~~~~~~~~~~~~~~~~~~~~~~~~~~~~~~~~~~~~~~~~~~~~~~~~~
\end{eqnarray}
 which are exact solution of the Einstein equations. The event horizons of the black hole are given by the singularities of the metric function which are the real roots of $\Delta_r=0$. The rang of $\theta$ depends strongly on the values of the NUT charge $n$, the rotational parameter $a$ and the cosmological constant $\Lambda =\frac{3}{l^2}$, taken here to be positive for dS and negative for AdS. When $l^{-2}=0$, the above metric reduces to \cite{8}
\begin{eqnarray}\label{3}
ds^{2}=-\frac{\Delta_r}{\rho^{2}}[dt+(2n\cos\theta-a\sin^{2}\theta) d\phi]^{2}~~~~~~~~~~~~\\
\nonumber
+\frac{\sin^{2}\theta}{\rho^{2}}[a dt-(r^{2}+n^{2}+a^{2})d\phi]^{2}+ \frac{\rho^{2}}{\Delta(r)}+\rho^{2} d\theta^{2}
    \end{eqnarray}
    where

\begin{equation}\label{4}
\rho^{2}=r^{2}+(n+a\cos(\theta))^{2}
\end{equation}

\begin{equation}\label{5}
\Delta_r= r^2-2 M r +a^{2}-n^{2}
\end{equation}
Now we consider a bulk massless scalar field $\Phi$ propagating in the background of (\ref{3}). The Klein-Gordon(KG) equation

\begin{eqnarray}\label{6}
\Box\phi=\frac{1}{\sqrt{-g}}\partial_{\mu}(\sqrt{-g}g^{\mu\nu}\partial_{\nu})\Phi=0
\end{eqnarray}
 can be simplified by assuming following form of the scalar field

\begin{eqnarray}\label{7}
  \Phi(t,r,\theta,\varphi)=\exp(-i\omega t+i m\varphi)S(\theta)  R(r)
\end{eqnarray}

The near region, which is the crucial region for demonstrating the origin of conformal structure, is defined by

\begin{eqnarray}\label{8}
 r\ll\frac{1}{\omega}~~~~~~~~~~~~~~~~M\ll\frac{1}{\omega}~~~~~~~~~~~~~~~~~~~n\ll\frac{1}{\omega}
\end{eqnarray}

Following in this background, the radial wave equation  reduces to following equation

\begin{eqnarray}\label{9}
\partial_r(\Delta\partial_r R)+[\frac{((r^2+n^2+a^2)\omega-ma)^2}{\Delta_r}+2ma\omega-l(l+1)]R=0
\end{eqnarray}
Let us consider the extreme Kerr-Bolt black hole. In this case the Hawking temperature is vanishing, $\Delta=(r-r_+)^2, r_+=M $ and $ M^2=a^2-n^2$. In the low frequency limit, we have the radial equation as

\begin{eqnarray}\label{10}
[\partial_r(\Delta\partial_r)+\frac{2(2M\omega)((r_+^2+a^2+n^2)\omega-ma)}{r-r_+}~~~~~~~~~~~~~~~~~~~~~~\\
\nonumber
+\frac{((r_+^2+a^2+n^2)\omega-ma)^2}{(r-r_+)^2}]R(r)=l(l+1)R(r)~~~~~~~~~~~~~~
\end{eqnarray}

\section{ Hidden conformal symmetry }
Following \cite{6} we now show that equation (\ref{10}) can be reproduced by the introduction of conformal coordinates.
 We will show that for massless scalar field $\Phi$, there exist a hidden $SL(2,R)$ conformal symmetry acting on the solution space. We can read out the left  temperature of dual conformal field theory (in the extremal limit we have only the left sector).
 We introduce the conformal coordinates

\begin{equation}\label{11}
\omega^{+}=\frac{1}{2}(\alpha_1 t+\beta_1 \varphi-\frac{\gamma_1}{r-r_+}),~~~~~~
\end{equation}

\begin{eqnarray}\label{12}
\omega^{-}=\frac{1}{2}(\exp(2\pi T_L\varphi+2n_Lt)-\frac{2}{\gamma_1})
\end{eqnarray}

\begin{eqnarray}\label{13}
 y=\sqrt{\frac{\gamma_1}{2(r-r_+)}}\exp(\pi T_L\varphi+n_L t)
\end{eqnarray}

 Now we can define the vector fields

\begin{eqnarray}\label{14}
H_1=i\partial_{+},~~~~
\end{eqnarray}

\begin{eqnarray}\label{15}
H_0=i(\omega^{+}\partial_{+}+\frac{1}{2}y\partial_{y}),~~~~
~
\end{eqnarray}

\begin{eqnarray}\label{16}
~H_{-1}=i((\omega^{+})^2\partial_{+}+\omega^{+}y\partial_{y}-y^2\partial_{-})~~~
\end{eqnarray}
and
\begin{eqnarray}\label{17}
\overline{H}_1=i\partial_{-}
\end{eqnarray}

\begin{eqnarray}\label{18}
\overline{H}_0=i(\omega^{-}\partial_{-}+\frac{1}{2}y\partial_{y})
\end{eqnarray}

\begin{eqnarray}\label{19}
\overline{H}_{-1}=i((\omega^{-})^2\partial_{-}+\omega^{-}y\partial_{y}-y^2\partial_{+})
\end{eqnarray}
Which each satisfy the $SL(2,R)$ algebra

\begin{eqnarray}\label{20}
 ~~[H_0,H_{\pm1}]=\mp i H_{\pm 1},~~~~~~~~[H_{-1},H_1]=-2iH_0
\end{eqnarray}
and

\begin{eqnarray}\label{21}
~[\overline{H}_0,\overline{H}_{\pm1}]=\mp i \overline{H}_{\pm 1},~~~~~~~~[\overline{H}_{-1},\overline{H}_1]=-2i\overline{H}_0
\end{eqnarray}

The quadratic Casimir is

\begin{eqnarray}\label{22}
H^2=\widetilde{H}^2=-H_{0}^2+\frac{1}{2}(H_1H_{-1}+H_{-1}H_{1})=\frac{1}{4}(y^2\partial_{y}^2-y\partial_{y})+y^2\partial_{+}\partial_{-}
\end{eqnarray}

The crucial observation is that these Casimir, when written in term of $ \varphi$ ,t and r reduces to the radial equation

\begin{eqnarray}\label{23}
H^2=\partial_r(\Delta\partial_r)-(\frac{\gamma_1(2\pi T_L\partial_t-2n_L\partial_{\varphi})}{A(r-r_+)})^2~~~~~~~~~~~~~~~~\\
\nonumber
-\frac{2\gamma_1(2\pi T_L\partial_t-2n_L\partial_{\varphi})}{A(r-r_+)}(\beta_1\partial_t-\alpha_1\partial_{\phi})~~~~~~~~~~~~~~
\end{eqnarray}
where $A=2\pi T_L\alpha_1-2n_L\beta_1$ and $\Delta=(r-r_+)^2$.
The equation (\ref{10}) could  be rewritten as the $SL(2,R)$ Casimir (\ref{23}) with the identification

\begin{eqnarray}\label{24}
\alpha_1=0~~~~~~~~~~~\beta_1=\frac{\gamma_1}{a}~~~~~~~~T_L=\frac{r_+^2+a^2+n^2}{4\pi a r_+}~~~~~~~~~~n_L=-\frac{1}{4M}
\end{eqnarray}
The identification of $T_L$ and $n_L$ are in consistence with existing result \cite{7}.

We can directly calculate all the $SL(2,R)$ generators in terms of black hole coordinates

\begin{eqnarray}\label{25}
H_1=\frac{2ia}{\gamma_1}(4a\partial_t+\partial_{\varphi})~~~~~~~~~~~~~~~~~~~~~
\end{eqnarray}

\begin{eqnarray}\label{26}
H_0=i(-(r-M)\partial_r+8a\phi\partial_t+2a\varphi\partial_{\varphi})~~~~~~~~~
\end{eqnarray}

\begin{eqnarray}\label{27}
H_{-1}=i(-\frac{\gamma_1\varphi}{a}(r-M)\partial_r+\frac{2M\gamma_1}{r-M}\partial_t~~~~~~~~~~~~~~~~~~~~~~~~~~\\
\nonumber
+\frac{a\gamma_1}{2}((\frac{\varphi}{a})^2+\frac{1}{(r-M)^2})(4a\partial_t+\partial_{\varphi}))~~~~~~~~~~~~~~~~~~~~~~~~~~
\end{eqnarray}

and

\begin{eqnarray}\label{28}
\overline{H}_1=\exp(-2\pi T_L\varphi-2n_L t)((r-M)\partial_r-(\frac{M}{2}+\frac{4a^2}{r-M})\partial_t-(\frac{a}{r-M})\partial_{\varphi})
\end{eqnarray}

\begin{eqnarray}\label{29}
\overline{H}_0=i(-\frac{2}{\gamma_1}\exp(-2\pi T_L\varphi-2n_L t)(r-M)\partial_r~~~~~~~~~~~~~~~~~~~~~~~~~~~~~~~~~~~~~~~~~\\
\nonumber
-\frac{M}{2}(1-\frac{2}{\gamma_1}\exp(-2\pi T_L\varphi-2n_L t))\partial_t~~~~~~~~~~~~~~~~~~~~~~~~~~~~~~~~~~~~~~~~~~~~~\\
\nonumber
-\frac{2a\exp(-2\pi T_L\varphi-2n_L t)}{\gamma_1(r-M)})(4a\partial_t+\partial_{\varphi})~~~~~~~~~~~~~~~~~~~~~~~~~~~~~~~~~~~~~~~~~~~~
\end{eqnarray}

\begin{eqnarray}\label{30}
\overline{H}_{-1}=i[-\frac{1}{2}(\exp(2\pi T_L\varphi+2n_L t)-\frac{4}{\gamma_1^2}\exp(-2\pi T_L\varphi-2n_L t))(r-M)\partial_r\\
\nonumber
-(\exp(2\pi T_L\varphi+2n_L t)-\frac{4}{\gamma_1}+\frac{4}{\gamma_1^2}\exp(-2\pi T_L\varphi-2n_L t))M\partial_t~~~~~~~~\\
\nonumber
-\frac{a}{2(r-M)}(\exp(2\pi T_L\varphi+2n_L t)+\frac{4}{\gamma_1^2}\exp(-2\pi T_L\varphi-2n_L t))(4a\partial_t+\partial_{\varphi})]
\end{eqnarray}
Actually there exist one degree of freedom $\gamma_1$ to define the vector fields, without affect the form of Casimir.

\section{Real-time correlator  }
The radial equation of the extremal Kerr-Bolt black holes take the form

\begin{eqnarray}\label{31}
H^2\Phi(r)=l(l+1)\Phi(r)
\end{eqnarray}

 where $l$ is a r-independent parameter contributing the conformal weight.
 With the ansatz (\ref{7}), the radial equation can be written as

\begin{eqnarray}\label{32}
[\partial_r(\Delta\partial_r)+\frac{B}{r-r_+}
+\frac{C^2}{(r-r_+)^2}]R(r)=l(l+1)R(r)~~~~~~~~~~~~~~
\end{eqnarray}
where

\begin{eqnarray}\label{33}
C=((r_+^2+a^2+n^2)\omega-ma)~~~~~~~~~~~\\
\nonumber
B=2(2M\omega)((r_+^2+a^2+n^2)\omega-ma)
\end{eqnarray}

Introducing $z=\frac{-2iC}{r-r_+}$, we get the equation

\begin{eqnarray}\label{34}
\frac{d^2R}{dz^2}+(\frac{\frac{1}{4}-m_s^2}{z^2}+\frac{k}{2}-\frac{1}{4})R(z)=0
\end{eqnarray}
where

\begin{eqnarray}\label{35}
k=i2M\omega,~~~~~~~~~~~~~m_s^2=\frac{1}{4}+l(l+1)
\end{eqnarray}
This equation has the solution

\begin{eqnarray}\label{36}
R(z)=C_1R_+(z)+C_2R_-(z)
\end{eqnarray}
where

\begin{eqnarray}\label{37}
R_{\pm}(z)=\exp(-\frac{z}{2})z^{\frac{1}{2}\pm m_s}F(\frac{1}{2}\pm m_s-k,1\pm2m_s,z)
\end{eqnarray}
are two linearly independent solution, $F$ is the Kummer function and could be expand in two limits. \\
1. Near horizon $r\rightarrow r_+$ so $z\rightarrow \infty$

\begin{eqnarray}\label{38}
F(\alpha.\gamma,z)\sim\frac{\Gamma(\gamma)}{\Gamma(\alpha)}\exp(-i\alpha\pi)z^{-\alpha}+\frac{\Gamma(\gamma)}{\Gamma(\alpha)}\exp(z)z^{\alpha-\gamma}
\end{eqnarray}
2. When $r$ goes asymptotically to infinity, $z\rightarrow 0$, $F\rightarrow 1$. In this case the solution has asymptotically behavior,

\begin{eqnarray}\label{39}
R\sim C_1r^{-h}+C_{2}r^{1-h}
\end{eqnarray}
where $h$ is the conformal weight

\begin{eqnarray}\label{40}
h=\frac{1}{2}+m_{s}=\frac{1}{2}+\sqrt{\frac{1}{4}+l(l+1)}
\end{eqnarray}
and

\begin{eqnarray}\label{41}
C_1=-\frac{\Gamma(1-2m_s)}{\Gamma(\frac{1}{2}-m_s-k)}D,~~~~~~~~~~C_2=\frac{\Gamma(1+2m_s)}{\Gamma(\frac{1}{2}+m_s-k)}D
\end{eqnarray}
where $D$ is a constant.
The retarded Green function should be read \cite{9}

\begin{eqnarray}\label{42}
G_R\sim\frac{C_1}{C_2}\propto\frac{\Gamma(1-2h)\Gamma(h-k)}{\Gamma(2h-1)\Gamma{(1-h-k)}}.
\end{eqnarray}

\section{Microscopic description}
In the extremal limit, the microscopic entropy comes the left sector and for Kerr-Bolt black holes $c_L=12Ma$, so

\begin{eqnarray}\label{43}
S=\frac{\pi^2}{3}c_LT_L=\pi(r_+^2+a^2+n^2)
\end{eqnarray}
in agreement with the \cite{7}.
 We can determine the conjugate charge from the first low of thermodynamics. We begin with nonzero $T_H$, then take limit to set it to zero.
 From the first low of thermodynamics,

\begin{eqnarray}\label{44}
\delta S=\frac{\delta M-\Omega_H\delta J -\phi \delta Q}{T_H}
\end{eqnarray}
we get

\begin{eqnarray}\label{45}
\delta S=2\pi(2M\delta M)+4\pi\frac{(2M^2+2n^2)\delta M-a\delta J}{2\sqrt{M^2-a^2+n^2}}
\end{eqnarray}

In the limit $T_H\rightarrow 0$ the second term turns to zero \cite{6},
 we identify

\begin{eqnarray}\label{46}
\delta M=\omega,~~~~~~~~~~~~~~~\delta E_L=\omega_L
\end{eqnarray}
with

\begin{eqnarray}\label{47}
\omega_L=\frac{(2M^2+2n^2)M}{J}\omega
\end{eqnarray}
then

\begin{eqnarray}\label{48}
\delta S=\frac{\delta E_L}{T_L}=\frac{\omega_L}{T_L}
\end{eqnarray}
Note that the identification (\ref{47}) is the same as the one found in the study of non-extremal Kerr-Bolt black holes. The retarded Green function in extremal Kerr-Bolt black holes should be rewritten as

\begin{eqnarray}\label{49}
G_R\sim\frac{\Gamma(1-2h)\Gamma(h-i2M\omega)}{\Gamma(2h-1)\Gamma(1-h-i2M\omega)}\\
\nonumber
\frac{\Gamma(1-2h)\Gamma(h-i\frac{\omega_L}{2\pi T_L})}{\Gamma(2h-1)\Gamma(1-h-i\frac{\omega_L}{2\pi T_L})}
\end{eqnarray}
In a two-dimensional conformal field theory, the two-point functions of the primary operators
are determined by the conformal invariance \cite{6}. The retarded correlator gives the Euclidean correlator with this relation:

\begin{eqnarray}\label{50}
G_E(\omega_{L,E})=G_{R}(i\omega_{L,E}),~~~~~~~~~~~~~~~~~\omega_{L,E}>0.
\end{eqnarray}
At finite temperature,  $\omega_{L,E}$  take discrete value of the Matsubara frequency

\begin{eqnarray}\label{51}
\omega_{L,E}=2\pi m_LT_L
\end{eqnarray}
The momentum space Euclidean correlator is given by \cite{6}:

\begin{eqnarray}\label{52}
G_E\sim T_{L}^{2h_L-1}\exp(i\frac{\omega_{L,E}}{2T_L})\Gamma(h_L+\frac{\omega_{L,E}}{2\pi T_L}))\Gamma(h_L-\frac{\omega_{L,E}}{2\pi T_L}))
\end{eqnarray}

The real-time correlator (\ref{49}) is obviously in agreement with the CFT prediction $\cite{6}$.

\section{Conclusion}
In this paper, we studied the hidden conformal symmetry of extremal Kerr-Bolt black holes. We solved the wave equation of a massless scalar field in the
background of extremal Kerr-Bolt spacetimes and find in the "near region", the wave equation can
be written in terms of the $SL(2,R)$ quadratic Casimir. In another term the quadratic Casimir when written in term of coordinates
 $\varphi$, $t$ and $r$ reduces to the radial equation of massless scalar field. In the previous paper \cite{7} we
considered the four-dimensional Kerr-Bolt spacetimes and have shown they have a hidden conformal
symmetry. In the extremal case we could not use the previous conformal coordinates as we have used in \cite{7}. In the extremal case the Hawking temperature is vanishing.
Therefore, only the left (right) temperature of the dual CFT is non-vanishing and
the excitations of the other sector are suppressed. In the present paper we have done our job by using the new conformal coordinate, where first time introduced in \cite{6}. So according to this results there exist dual CFT description of extremal Kerr-Bolt black holes. Also the macroscopic entropy
and the absorption cross section of a near-region scalar field match precisely to that of
microscopic dual CFT side.
\section{Acknowledgement}
The authors would like to thank Bin Chen for reading the manuscript.

\end{document}